\documentclass[12pt]{article}

\usepackage{amsmath,amsfonts,amssymb,amsthm,amstext,eucal,
amsbsy,amsopn,amsgen}

\newtheorem{thm}{Theorem}[section]

\theoremstyle{definition}
\newtheorem{exa}[thm]{Example}

\theoremstyle{remark}

\numberwithin{equation}{section}

\newcommand{\al}{\alpha}
\newcommand{\be}{\beta}
\newcommand{\ga}{\gamma}
\newcommand{\de}{\delta}
\newcommand{\la}{\lambda}
\newcommand{\om}{\omega}
\newcommand{\si}{\sigma}
\newcommand{\De}{\Delta}
\newcommand{\Om}{\Omega}
\newcommand{\Real}{\mathbb R}
\newcommand{\eps}{\varepsilon}
\newcommand{\yh}{\hat{y}}
\newcommand{\xh}{\hat{x}}
\newcommand{\qh}{\hat{q}}
\newcommand{\ph}{\hat{p}}
\newcommand{\Lh}{\hat{L}}
\newcommand{\Yh}{\hat{Y}}
\newcommand{\Qh}{\hat{Q}}
\newcommand{\erb}{\mathbf{e_r}}
\newcommand{\etb}{\mathbf{e_{\theta}}}
\newcommand{\epb}{\mathbf{e_{\phi}}}
\newcommand{\qb}{\mathbf{q}}
\newcommand{\pb}{\mathbf{p}}
\newcommand{\Lb}{\mathbf{L}}
\newcommand{\Lhb}{\mathbf{\hat{L}}}
\newcommand{\Imp}{\Rightarrow}
\newcommand{\ra}{\rightarrow}
\newcommand{\To}{\longrightarrow}

\newcommand{\M}{\mathcal{M}}

\newcommand{\bx}{\bar{x}}
\newcommand{\tx}{\tilde{x}}
\newcommand{\dx}{\dot{x}}
\newcommand{\bq}{\bar{q}}
\newcommand{\bp}{\bar{p}}
\newcommand{\pa}{\partial}
\newcommand{\pl}{\overset{\leftarrow}{\partial}}
\newcommand{\pr}{\overset{\rightarrow}{\partial}}
\newcommand{\dl}{\overset{\leftarrow}{\delta}}

\begin{document}

\title{Exploring the second class constraint quantization approach
proposed by Batalin and Marnelius}

\author{{\bf Michael Chesterman\footnote{\texttt{M.J.Chesterman@qmul.ac.uk}}}\\ {\small The Physics
Department,}\\{\small Queen Mary,}\\ \small{University of London,}
\\ {\small Mile End Road, E1 4NS, UK}}


\date{January 2002\\{\small Preprint QMUL-PH-02-02}}

\maketitle

\begin{abstract}
I extend upon the paper by Batalin and Marnelius, in which they
show how to construct and quantize a gauge theory from a
Hamiltonian system with second class constraints. Among
the avenues explored, their technique is analyzed in relation to
other well-known methods of quantization and a bracket is defined,
such that the operator formalism can be fully developed. I also
extend to systems with mixed class constraints and look at some
simple examples.
\end{abstract}

\section{Introduction}

The subject of this work concerns a paper \cite{Batalin:2001hs} by
Batalin and Marnelius, in which they propose a novel way of
constructing a gauge theory from a Hamiltonian system with second
class constraints. Their work extends on ideas first developed in
\cite{Lyakhovich:2001cm}.

The covariant quantization of general Hamiltonian systems, with
second class constraints, is no easy task. Two of the best methods
so far are `constraint conversion'
\cite{Batalin:1987fm,Batalin:1990mb,Batalin:1991jm} and `gauge
unfixing' \cite{Harada:1990aj,Mitra:1990mp}. The first
introduces additional variables in such a way that the second class
constraints are converted into first class ones. The second on the
other hand, finds a subset of half of the constraints in involution
with each other, and then discards the remaining constraints as
`gauge fixing' conditions, thus leaving a first class constraint
surface. The resultant gauge theories can be quantized using the
full power of Hamiltonian BRST (BFV) quantization. However, even
the above two methods are limited. For example, `gauge unfixing' is
only possible locally in general.

Although the authors of the subject paper \cite{Batalin:2001hs}
constructed the correct path integral following the superfield
version of antifield (BV) quantization, there were some stones left
unturned. In particular, an operator formalism was not developed
and the question of the position of the technique, in relation
both to standard Dirac bracket quantization and the above two
approaches, was not fully explored. In the work which follows, I
explore these points and others in some detail and also extend
to systems with mixed class constraints.

I do not intend to review the subject paper in any depth, as this
would be superfluous. Rather, this work is designed to be read in
conjunction with the subject paper. However, all equations and
identities from the subject paper that I use are also shown here,
either in section \ref{sec:BMequations}, or as needed in the text.
To avoid confusion, I also use the same notation.

\section{Some Preliminaries}
\label{sec:BMequations}

\subsection{The Setting}
\label{sec:Setting}

We start, as in \cite{Batalin:2001hs}, with symplectic
supermanifold $\M$ which has co-ordinates $x^i$, $i=1,\ldots,2N$
with $\eps^i=\eps(x^i)$. The non-degenerate symplectic two-form
$\om_{ij}(x)$ is closed
\begin{equation}\label{eq:omclosed}
d\om=0 \Leftrightarrow \pa_i\om_{jk}(x)(-1)^{(\eps^i+1)\eps^k} +
cycle(i,j,k)=0
\end{equation}
and has properties
\begin{eqnarray}
\om_{ij}(x)=\om_{ji}(x)(-1)^{(\eps^i+1)(\eps^j+1)} &&
\eps(\om_{ij})=\eps^i + \eps^j.
\end{eqnarray}
Since $\om_{ij}$ is non-degenerate, there exists an inverse
$\om^{ij}(x)$, in terms of which the Poisson bracket is defined by
\begin{equation}\label{eq:Poissonbracket}
\{A(x),B(x)\}=A(x)\pl_i \om^{ij}(x) \pr_j B(x).
\end{equation}
$w^{ij}$ obeys
\begin{eqnarray}
\om^{ij}(x)=-\om^{ij}(x)(-1)^{\eps^i\eps^j} &&
\eps(\om^{ij})=\eps^i + \eps^j.
\end{eqnarray}
The Poisson bracket \eqref{eq:Poissonbracket} satisfies the Jacobi
identities since \eqref{eq:omclosed} implies
\begin{equation}
\om^{ij}\pl_l \om^{lk}(-1)^{\eps^i\eps^k} + cycle(i,j,k)=0.
\end{equation}

We consider a Hamiltonian $H(x)$, with $2M<2N$ irreducible second
class constraints $\theta^{\al}(x)$, which satisfy regularity
condition
\begin{equation}\label{eq:regular}
\text{Rank } \theta^{\al}(x){\frac{\pl}{\partial x^i}}
|_{\theta=0}=2M
\end{equation}
and class condition
\begin{equation}\label{eq:classcnd}
\text{Rank }{\{\theta^{\al}(x),\theta^{\be}(x)\}}|_{\theta=0}=2M.
\end{equation}
Constraints $\theta^{\al}=0$ determine a constraint surface
$\Gamma$, which is itself a symplectic supermanifold of dimension
$(2N-2M)$.

The equations of motion can be calculated from the extended
Hamiltonian form of the action
\begin{equation}
S_E[x]=\int{dt\, x^i \bar{\om}_{ij}(x) \dx^j - H(x)
-\la_{\al}\theta^{\al}(x)},
\end{equation}
where \cite{Batalin:1998pz}
\begin{equation}\label{eq:ombar}
\bar{\om}_{ij}(x)=\int_{0}^{1}{\om_{ij}(\al x)\al \, d\al} \Imp
\om_{ij}(x)=(x^k\pr_k+2)\bar{\om}_{ij}(x).
\end{equation}
Equations \eqref{eq:omclosed} and \eqref{eq:ombar} yield the
required property that
\begin{equation}
\de \int{dt \, x^i\bar{\om}_{ij}(x)\dx^j}=\int{dt\, \de x^i
\om_{ij}(x)\dx^j}.
\end{equation}
In general, $x^i$ will be canonical co-ordinates with constant
$\om^{ij}$ and $\bar{\om}^{ij}=\frac{1}{2} \om^{ij}$.

In the subject paper, the authors parameterize the constraint
surface $\Gamma$ with the original phase space co-ordinates $x^i$,
constructing functions $\bx^i(x)$ satisfying
\begin{eqnarray}\label{eq:xbar1}
\theta^{\al}(\bx^i(x))=0\label{eq:map1}\\\label{eq:xbar2}
\bx^i(\tx)=\tx^i \label{eq:map2} && \text{for } \tx^i \in \Gamma.
\end{eqnarray}
As there are $2M$ more co-ordinates than the dimension of $\Gamma$,
there are $2M$ gauge symmetries denoted by \eqref{eq:gaugegen}. The
BM gauge invariant action is
\begin{equation}\label{eq:BMaction}
S[\bx(x)]=\int{dt\, \bx^i \bar{\om}_{ij}(\bx) \dot{\bx}^j -
H(\bx)}.
\end{equation}

\subsection{Equations and Identities from the BM paper}
\label{sec:BMidentities}

\begin{equation}\label{eq:Zdef}
\theta^{\al}(x)\pl_i Z^i_{\be}(x)=\delta^{\al}_{\be}
\end{equation}
\begin{eqnarray}
P^i_j(x)=\de^i_j-Z^i_{\al}(x)(\theta^{\al}(x)\pl_j)\label{eq:P}\\
P^i_j(x)P^j_k(x)=P^i_k(x)\\
P^i_j(x)Z^j_{\al}(x)=0 \label{eq:PZ}
\end{eqnarray}
\begin{equation}\label{eq:xtilda}
\frac{d\tx^i}{d\la}=Z^i_{\al}(\tx)\theta^{\al}(\tx)
\end{equation}
\begin{equation}\label{eq:dxbar}
\bx^i(x)\pl_k=P^i_m(\bx)\sigma^m_k(x)
\end{equation}
\begin{eqnarray}
\si^m_k(\bx)=\de^i_k && \det \si^i_k(x)\neq 0
\end{eqnarray}
\begin{equation}\label{eq:si_ambiguity}
\si^m_k(x) \To \si^m_k(x)+Z^m_\al(\bx)M^\al_k(x)
\end{equation}

\begin{eqnarray}\label{eq:gaugegen}
\bx^i(x) \pl_k G^k_{\al}(x)=0 && G^k_{\al}(x)=(\si^{-1})^k_m
Z^m_{\al}(\bx(x))
\end{eqnarray}

\section{Conditions for existence of continuous function $\bx(x)$}

In sections 3 and 4 of the subject paper\cite{Batalin:2001hs}, the
authors argue that given second class constraint functions
$\theta^{\al}(x)$ obeying regularity conditions \eqref{eq:regular}
and \eqref{eq:classcnd}, the continuous function $\bx^i(x)$ always
exists. In fact, a small modification of their argument is
necessary. The mathematics required to cast light on this point is
well known \cite{Nakahara:1990th}. Consider a generic constraint
surface $\Gamma$, a submanifold of $\M$. If a continuous function
$\bx: \M \rightarrow \Gamma$ exists, obeying \eqref{eq:map1} and
\eqref{eq:map2}, then $\bx$ is said to be a retraction and
$\Gamma$ a retract of $\M$. Furthermore, if there exists a
continuous map $H:\M\times I \rightarrow \M$, where $I$ is the
interval $[0,1]$, such that \begin{eqnarray}\label{eq:dretract1}
H(x,0)=x & H(x,1) \in \Gamma & \text{for any } x \in \M\\
H(x,s)=x & \text{for any } x \in \Gamma &\text{and any } s \in
I\label{eq:dretract2}
\end{eqnarray}
then $\Gamma$ is said to be a deformation retract of $\M$. Not all
retracts are deformation retracts, however, it's reasonable here
to just check for a deformation retract with $H(x,1)=\bx(x)$. The
properties \eqref{eq:dretract1} and \eqref{eq:dretract2} of
$H(x,s)$ imply that the identity function on $\M$ is homotopic to
the function $\bx$. Thus $\M$ and $\Gamma$ have the same homotopy
type. In other words, $\M$ and $\Gamma$ must have the same
fundamental group:
\begin{equation}\label{eq:fundgrp}
\pi_1(\M)=\pi_1(\Gamma).
\end{equation}
This topological condition can be translated into a restriction on
the constraint functions. Batalin and Marnelius explicitly
construct $\bx^i(x)$ as a deformation retract, via the
differential equation \eqref{eq:xtilda} for continuous function
$\tx^i(\la,x)$ in section 4 of the subject paper. Given initial
conditions $\tx^i(0,x)=x^i$, the authors define
\begin{equation}
\bx^i(x)=\lim_{\la \ra -\infty} \tx^i(\la,x).
\end{equation}
By construction, this definition obeys both \eqref{eq:map1} and
\eqref{eq:map2}. The differential equation \eqref{eq:xtilda}
uniquely dictates the path in $\M$ that the point $x^i$ takes to
$\bx^i(x)$ which is traced by $\tx^i(\la,x)$ over the interval
$\la=[0,-\infty]$. We can see that $\tx(\lambda,x)$ is a homotopy
as defined in \eqref{eq:dretract1} and \eqref{eq:dretract2} if we
substitute $s=1-e^{\lambda}$, since
\begin{eqnarray}
\tx(s=0,x)=x && \tx(s=1,x)=\bx(x).
\end{eqnarray}
The only problem with the Batalin Marnelius definition of $\bx(x)$
is that $Z^i_{\al}(x)$ can be singular off the constraint surface
in the general case. The requirement \eqref{eq:regular} on the
matrix $\theta^{\al}(x)\pl_i$ only applies on the constraint
surface. Away from the surface, $\text{Rank
}(\theta^{\al}(x)\pl_i) \leq 2M$, but $Z^i_{\al}(x)$ obeying
\eqref{eq:Zdef} can only be defined when
\begin{eqnarray}\label{eq:Zsmooth}
\text{Rank}\;\theta^{\al}(x)\pl_i = 2M & \text{for any } x \in \M.
\end{eqnarray}
The above equation is actually equivalent to \eqref{eq:fundgrp}.
So long as $Z^i_{\al}(x)$ can be smoothly defined, the
differential equation for $\tx(\la,x)$ leads to a unique and
smooth function $\bx(x)$ in terms of $Z^i_{\al}(x)$ and
$\theta^{\al}(x)$. The points $x \in \M$ where \eqref{eq:Zsmooth}
doesn't hold correspond to singularities in $Z^i_{\al}(x)$ and
discontinuities or singularities in $\bx(x)$. In order to
illustrate the connection  between \eqref{eq:fundgrp} and
\eqref{eq:Zsmooth}, consider a simple example. Suppose
$\M=\Real^2\times\Real^2$, with canonical co-ordinates $(q^i,
p_i)$, $i=1,2$. Also, suppose there's one first class constraint
$\theta=(q^1)^2+(q^2)^2-R^2=0$ where $R$ is a constant. The class
of $\theta$ isn't important. The constraint surface
$\Gamma=S^1\times\Real^2$. If we forget the co-ordinates
$(p_1,p_2)$ for now, clearly $\pi_1(S^1) \neq\pi_1(\Real^2)$ and
one can expect a singular point in $\bx(x)$ anywhere inside the
circle, depending on the definition of $\bx(x)$. In fact, the
singular point is determined by $\theta$.
$\text{Rank}\;\theta(x)\pl_i=1$ everywhere, except at singular
point $q^1=q^2=0$. The solution is to exclude $q^1=q^2=0$ from
$\M$, then $\pi_1(\Real^2-\{0\})=\pi_1(S^1)$.

One should ask how severe the restriction \eqref{eq:Zsmooth} is.
In order to use the Dirac bracket for second class constraints
$\theta^{\al}(x)$ , the matrix $\{\theta^{\al},\theta^{\be}\}$
must be invertible everywhere in $\M$, i.e.
\begin{eqnarray}\label{eq:thetarank1} \text{Rank
}{\{\theta^{\al}(x),\theta^{\be}(x)\}}= 2M & \text{for any } x \in
\M.
\end{eqnarray}
Now since
\begin{equation}\label{eq:thetarank2}
\{\theta^{\al},\theta^{\be}\}=\theta^{\al}\pl_i \om^{ij} \pr_j
\theta^{\be},
\end{equation}
we can write an inequality for the rank of the matrix
$\theta^{\al}\pl_i$ using the following result. For $N\times N$
matrices $A_i$ $i=1,\ldots, n$ and $B$ where
\begin{equation}
A_1 A_2 \ldots A_n =B,
\end{equation}
it follows that
\begin{equation}
\max{(\text{Ker } A_1, \ldots , \text{Ker } A_n)} \leq \text{Ker }
B \leq \sum_{i=1}^{n}{\text{Ker } A_i},
\end{equation}
where $\text{Ker } A_i=N- \text{Rank } A_i$. All matrices in
\eqref{eq:thetarank2} can be made square $2N\times 2N$ by filling
with zeros as necessary. The resultant inequality is
\begin{equation}
\text{Rank } \theta^{\al}\pl_i \geq \text{Rank }
\{\theta^{\al},\theta^{\be}\} \geq \text{Rank }
\theta^{\al}\pl_i-\text{Ker }\theta^{\al}\pl_i,
\end{equation}
thus \eqref{eq:thetarank1} implies \eqref{eq:Zsmooth}. The BM
technique can be used for all second class systems where a Dirac
bracket can be smoothly defined. Usually $\M \equiv R^n \times
R^n$ which has trivial fundamental group $\pi_1(\M)$. This might
seem to severely limit the possible constraint surfaces that can
be dealt with. However, exclusion of strategic point(s), or in
general, hyperplanes from $\M$, may sometimes be possible, with
drastic effects on $\pi_1(\M)$. These points are ones where
\eqref{eq:Zsmooth} doesn't hold. This is dynamically allowed if
the excluded points are fixed points.

\section{A suitable bracket on $\M$}\label{sec:Mybracket}

In a previous paper \cite{Lyakhovich:2001cm}, a third condition was
placed on $\bx^i(x)$
\begin{equation}\label{eq:Poisson}
{\{\bx^i(x), \bx^j(x)\}}={\{x^i,x^j\}}_D|_{\:x \ra \bx(x)},
\end{equation}
where $\{,\}$ and $\{,\}_D$ are the Poisson and Dirac bracket
respectively on $\M$. In the subject paper \cite{Batalin:2001hs},
condition \eqref{eq:Poisson} was considered to restrict the choice
of gauge theory and was removed. In this spirit, one can instead
search for a new bracket $\{,\}_M$ on $\M$ with the property
\begin{equation}\label{eq:mybracket}
\{A(\bx(x)),B(\bx(x))\}_M=\{A(x),B(x)\}_D|_{x \ra \bx(x)},
\end{equation}
where $A(\bx(x))$ and $B(\bx(x))$ are arbitrary, gauge invariant
observables. So
\begin{equation} \label{eq:Mbracket}
\{\bx^i(x),\bx^j(x)\}_M=\{x^i,x^j\}_D|_{x \ra \bx(x)}.
\end{equation}
Utilizing \eqref{eq:dxbar}, equation \eqref{eq:Mbracket} may be
written as
\begin{equation}
P^i_m(\bx)\si^m_k(x)\om^{kl}_M(x)\si_l^n(x)P_n^j(\bx)=\om^{ij}_D(\bx),
\end{equation}
where we define
\begin{equation}
\{a(x),b(x)\}_M=a(x)\pl_i\om^{ij}_M(x)\pr_j b(x)
\end{equation}
and similarly for the Dirac bracket which is characterized by
\begin{equation}
\om_D^{ij}(x)=\om^{ij}(x) - \{x^i,\theta^\al\}
\{\theta^\al,\theta^\be\}^{-1} \{\theta^\be ,x^j\}.
\end{equation}
Note that $\text{Rank }P^i_m(\bx)=2N-2M$, so $\om_M^{ij}(x)$ isn't
uniquely defined and furthermore this is also reflected by the
ambiguity in choosing $\si^m_k(x)$ shown in
\eqref{eq:si_ambiguity}. Firstly, a particular solution for
$\om^{ij}_M(x)$ must be found. Since $\{\theta^\al,x^j\}_D=0$ and
making use of the explicit expression for $P^i_j(x)$ in
\eqref{eq:P}, \begin{equation}\label{eq:PomP}
P^i_k(x)\om^{kl}_D(x)P_l^j(x)=\om^{ij}_D(x).
\end{equation}
It soon becomes clear that a particular solution is
\begin{equation}\label{eq:omM}
\om^{ij}_M(x)=(\si^{-1})^i_k(x)\om^{kl}_D(\bx)(\si^{-1})_l^j(x).
\end{equation}
In order to obtain the general solution for $\om^{ij}_M$ we add a
term $\De^{ij}$ such that
\begin{equation}
P^i_k(\bx)\De^{kl}_{\si}(x)P_l^j(\bx)=0,
\end{equation}
where $\De_{\si}^{kl}=(\si\De\si)^{kl}$ and where we also require
that $\De^{ij}=(-)(-)^{\eps^i\eps^j}\De^{ji}$. The $2M$ vectors
$Z_{\al}^i(x)$ form a basis for the kernel of the matrix
$P^i_k(x)$ as can be seen from \eqref{eq:PZ}. The most general
such expression is
\begin{equation}
\De_{\si}^{kl}(x)=Z^k_{\al}(\bx)J^{\al l}(x)
-(-1)^{\eps^{al}(1+\eps^k + \eps^l)} J^{\al k}(x)Z^l_{\al}(\bx),
\end{equation}
where $J^{\al k}(x)$ is some arbitrary expression such that
$\eps(J^{\al k})=\eps^k-\eps^{\al}$. Note that freedom in the
choice of $\De^{ij}(x)$ means freedom in the choice of $\text{Rank
} \om_M^{ij}$. One can have $\text{Rank } \om_M^{ij}=2N-2M$ as in
\eqref{eq:omM}, or choose $\De^{ij}(x)$ such that $\text{Rank }
\om_M^{ij}=2N$.

The bracket $\{,\}_M$ should also satisfy the Jacobi identity. If
only gauge invariant observables of the form $A(\bx(x))$ are ever
used inside the bracket we simply require that
\begin{equation}\label{eq:gaugeJacobi}
\begin{split}
&\{\{A(\bx),B(\bx)\}_M,C(\bx)\}_M\\
&\text{  }+(-)^{\eps^A(\eps^B+\eps^C)}\{\{B(\bx),C(\bx)\}_M,A(\bx)\}_M \\
&\text{  }
+(-)^{\eps^C(\eps^A+\eps^B)}\{\{C(\bx),A(\bx)\}_M,B(\bx)\}_M=0.
\end{split}
\end{equation}
Since $P^i_k(\bx)\om_D^{kj}(\bx)=\om_D^{ij}(\bx)$, one can
carefully show that
\begin{equation}
\{\{A(\bx),B(\bx)\}_M,C(\bx)\}_M=\{\{A(x),B(x)\}_D,C(x)\}_D|_{x \ra
\bx}.
\end{equation}
As the Dirac bracket obeys the Jacobi identity, so
\eqref{eq:gaugeJacobi} is true trivially. However, if one wants to
allow general functions inside the bracket too, the Jacobi
condition is considerably harder to solve and a full solution
won't be necessary here.

One notable property of $\{,\}_M$ is
\begin{equation}
\{A(\bx),B(\bx)\}_M \pl_i G^i_{\al}(x)=0,
\end{equation}
since $\{A(\bx),B(\bx)\}_M$ is a function of $\bx$ as can be seen
in \eqref{eq:mybracket}.

The bracket $\{,\}_M$ actually originates from the uniquely
defined, degenerate, symplectic two-form
\begin{equation}\label{eq:omMform}
{\om_M}_{ij}(x)=(\pr_{i}\bx^k) \,\om_{kl}(\bx(x))\,
(\bx^l\pl_{j}).
\end{equation}
One can see for example that \eqref{eq:BMaction} defines equations
of motion
\begin{equation}
{\om_M}_{ij}(x)\dx^j=\pr_i H(\bx(x))
\end{equation}
The fact that ${\om_M}_{ij}$ is degenerate means that $\dx^j$
isn't unique and we can check, using the identities in section
\ref{sec:BMidentities} and the explicit expression for the Dirac
bracket, that
\begin{equation}
\dx^j=\{x^j,H(\bx(x))\}_M
\end{equation}
defines the general solution.

\section{A review of the difficulties with the Dirac bracket}
\label{sec:Diracbracket}

It will be useful to review the explicit reasons why, in general,
the Dirac bracket quantization method fails \cite{Henneaux:1992ig}.
There are various ways to attempt operator quantization of the
system described in section \ref{sec:Setting}. One way is to find a
quantum representation of the Dirac bracket, with second class
constraints imposed as operator equations.
Another, possibly simpler, approach is to get rid of the
constraint conditions in the classical system before quantization,
by parameterizing the constraint surface $\Gamma$ with $2N-2M$
independent co-ordinates:
\begin{eqnarray}
y^{\mu} && \eps(y^\mu)=\eps^\mu,
\end{eqnarray}
where there exists a one-to-one continuous mapping $\bx^i(y)$,
obeying
\begin{equation}
\theta^{\al}(\bx(y))=0.
\end{equation}
The non-degenerate, induced, symplectic two-form is
\begin{equation}\label{eq:omy}
\om_{\mu\nu}(y)=(\pr_{\mu}\bx^i) \,\om_{ij}(\bx(y))\,
(\bx^j\pl_{\nu}).
\end{equation}
All observables $f(y)$ can be written in the form $F(\bx(y)$. We
now have a system with no constraints, action $S_E[\bx(y)]$ and in
general, non-trivial Poisson bracket constructed with
$\om^{\mu\nu}(y)$, the inverse of $\om_{\mu\nu}(y)$. It's well
known \cite{Henneaux:1992ig} that
\begin{equation}\label{eq:starbracket}
\{f(y),g(y)\}_* = \{F(x),G(x)\}_D |_{x\ra \bx(y)}
\end{equation}
for
\begin{eqnarray}
f(y)=F(\bx(y)) && g(y)=G(\bx(y))
\end{eqnarray}
\begin{equation}
\{f(y),g(y)\}_*=f(y)\pl_\mu \om^{\mu\nu}(y) \pr_\nu g(y).
\end{equation}
We now look for a consistent representation of the bracket
\eqref{eq:yrep} below, for independent operators $\hat{y}^\mu$:
\begin{equation}\label{eq:yrep}
[\yh^\mu,\yh^\nu]=i\hbar \om^{\mu\nu}(\yh)
\end{equation}
where
\begin{equation}
[\yh^\mu,\yh^\nu]=\yh^\mu \yh^\nu -(-)^{\eps^\mu \eps^\nu} \yh^\nu
\yh^\mu.
\end{equation}
The problem is that no such consistent representation exists in
general, except in particular cases, e.g. when $\om^{\mu\nu}$ is
constant or where $\yh^\mu$ form an algebra in the bracket $[,]$.

The situation is no easier from the path integral approach. One
can write the path integral in terms of $y^\mu$
\begin{equation}\label{eq:ypathint}
\int{[Dy^\mu]\; \text{sdet}(\om_{\mu\nu})^{1/2}
\exp{\frac{i}{\hbar}\int{dt \, S_E[\bx(y)]}}},
\end{equation}
or in terms of the original co-ordinates $x^i$,
\begin{equation}\label{eq:xpathint}
\int{[Dx^i] [D\la^{\al}]\; \text{sdet}(\om_{ij})^{-1/2}
\text{sdet}(\{\theta^\al,\theta^\be\})^{1/2}
\exp{\frac{i}{\hbar}\int{dt \, S_E[x]}}}.
\end{equation}
Both expressions may look deceptively simple. However, in the
co-ordinate representation for example, the problem is in the
boundary conditions. In the co-ordinate representation, both
\eqref{eq:ypathint} and \eqref{eq:xpathint} are quantum amplitudes
between states determined by the initial and final configuration
of fields in the path integral. We need to find a complete set of
$N-M$ independent, commuting operators $\hat{Q}^{\ga}$, in order
to label states.

The problem of finding global, commuting co-ordinates
$\hat{Q}^{\ga}$ is generally unsolvable and seems to be related to
the difficulty in the operator formalism of finding a consistent
representation of the bracket in \eqref{eq:yrep}. This explains why
the methods mentioned in the introduction to this paper are more
powerful, as they work with the original Poisson bracket
$\om^{ij}$, which furnishes a representation when, as is most often
the case, $x^i$ are canonical co-ordinates.

\section{Exploring the Batalin Marnelius gauge theory}
\label{sec:exploreBM}

It's true that two of the most effective approaches,
\cite{Batalin:1987fm,Batalin:1990mb,Batalin:1991jm} and
\cite{Harada:1990aj,Mitra:1990mp} described in the introduction,
to the covariant quantization of Hamiltonian systems with
irreducible second class constraints are by converting to an
equivalent gauge theory. However, its not easy to covariantly
quantize all gauge theories. We should therefore be specific about
which gauge theories are desirable. To this purpose, I define two
terms:

I define the `extended Hamiltonian form of the action' to be
\begin{equation} S_E[x]=\int{dt\, x^i \bar{\om}_{ij}(x) \dx^j -
H(x) -\la_{\al}\theta^{\al}(x) -\la_a g^a(x)},
\end{equation}
where $\bar{\om}_{ij}(x)$ is defined in \eqref{eq:ombar} and
$\theta^{\al}(x)$ and $g^a(x)$ are the irreducible second and first
class constraints respectively.

I define the action to be `desirable' when it can be written in the
form of $S_E[x]$ such that there are no $\theta^{\al}(x)$ terms and
where $\om_{ij}$ is constant and thus equal to $2\bar{\om}_{ij}$.
Usually, constant $\om_{ij}$ means that $S_E[x]$ is written in
terms of canonical co-ordinates $(q^i,p_i)$.

Note that we can get rid of the $\theta^{\al}(x)$ terms in
$S_E[x]$ by parameterizing the second class constraint surface
with non-degenerate co-ordinates $y^\mu$, as in section
\ref{sec:Diracbracket}. The cost is that $S_E[y]$ will then have
in general a symplectic two-form $\om_{\mu\nu}(y)$, as defined in
\eqref{eq:omy}, which depends on $y^\mu$. The two previously
mentioned methods convert the original action $S_E[x]$, assumed to
be in canonical co-ordinates, to a `desirable' one. `Desirable'
actions can be elegantly covariantly quantized using one of the
equivalent antifield (BV) or Hamiltonian BRST (BFV) approaches
\cite{Henneaux:1992ig}. As mentioned in section
\ref{sec:Diracbracket}, the system can be quantized for some
non-constant $\om_{ij}(x)$'s, but usually, when converting to a
gauge theory, we really mean a `desirable' gauge theory.

The point is that the Batalin Marnelius action, $S[\bx(x)]$, isn't
in `extended Hamiltonian form' because $\om_M(x)_{km}$, defined in
\eqref{eq:omMform}, is degenerate. It's thus not immediately
clear whether it's a `desirable' gauge theory or not.

One can convert $S[\bx(x)]$ to standard $S_E$ form, by treating
$x^i$ as ordinary Lagrangian co-ordinates and then following the
standard procedure for formation of the Hamiltonian etc. Assume
for simplicity that the system is bosonic and $\om_{ij}$ is
constant, as is most often the case, and introduce canonical
momenta and Poisson bracket $\{,\}_P$ as follows:
\begin{eqnarray}
p_{x^i}=\frac{\pa}{\pa\dx^i}L(x,\dx) \\
\{x^i,x^j\}_P=\{p_{x^i},p_{x^j}\}_P=0 && \{x^i,p_{x^j}\}_P=\de^i_j.
\end{eqnarray}
The Hamiltonian is
\begin{equation}
H(x,p_x)=H(\bx(x))
\end{equation}
and there are $2N$ irreducible, primary constraints:
\begin{eqnarray}
\phi_k=p_{x^k}-\frac{1}{2}\bx^i(x)\om_{ij}\pa_k\bx^j &&
\{\phi_k,\phi_m\}_P=\om_M(x)_{km},
\end{eqnarray}
with no further constraints obtained from consistency conditions
\begin{equation}
\dot{\phi_k}\approx \{\phi_k,H\}_P + v^m\{\phi_k,\phi_m\}_P \approx
0,
\end{equation}
where $A\approx B$ means $A=B$ on the constraint surface
$\phi_k=0$. Now, $\text{Rank }\{\phi_k,\phi_m\}_P=2N-2M$ and 2M
first class constraints can be projected out
\begin{eqnarray}
\phi_{\al}=\phi_k G^k_{\al}(x) &&
\{\phi_k,\phi_{\al}\}_P \approx 0.
\end{eqnarray}
As expected, $\phi_{\al}$ generate the gauge transformations
\begin{eqnarray}
\{x^i,\phi_{\al}\}_P=G^i_{\al}(x) &&
\{p_{x^i},\phi_{\al}\}_P=-p_{x^k}\pa_iG^k_{\al}(x)
\end{eqnarray}
and the action is
\begin{equation}
S_E[x,p_x]=\int{dt\, p_{x^i}\dx^i-H(\bx(x))-\la^k \phi_k(x,p_x)}.
\end{equation}
We now parameterize the second class constraint surface using
co-ordinates $(p_\al,x^i)$,
\begin{eqnarray}
x^k(p_\al,x^i)&=& x^k\\
p_{x^k}(p_\al,x^i) &=& \frac{1}{2}\bx^i\om_{ij}\pa_k\bx^j +
p_{\al}(\pa_m \theta^{\al})(\bx) \si^m_k(x).
\end{eqnarray}
So,
\begin{eqnarray}
\phi_k(p_\al,x^i)&=& p_{\al}(\pa_m \theta^{\al})(\bx) \si^m_k(x)\\
\phi_\al(p_\al,x^i)&=& p_\al,
\end{eqnarray}
using some of the identities in section \ref{sec:BMidentities}.
Note that on the first class constraints surface,
$\phi_k(p_\al,x^i)=0$ as required. We now calculate the
corresponding symplectic two-form in these co-ordinates,
\begin{equation}
\om_{(p_\al,x^i),(p_\be,x^j)}= \frac{\pa X^k}{\pa(p_\al,x^i)}
\Om_{km} \frac{\pa X^m}{\pa(p_\be,x^j)},
\end{equation}
where $X^k=(x^i,p_{x^i})$ and $\Om_{km}$ is the symplectic two-form
corresponding to the bracket $\{,\}_P$.
\begin{equation}
\om_{(p_\al,x^i),(p_\be,x^j)}=
\begin{pmatrix}
0 & \vdots &(\pa_m \theta^{\al})(\bx) \si^m_j(x) \\
\hdotsfor{3}\\
-\si^m_i(x)(\pa_m \theta^{\be})(\bx) &\vdots & {\om_M}_{ij}(x)+
p_{\al}A^{\al}_{ij}(x)
\end{pmatrix}_{(p_\al,x^i),(p_\be,x^j)}
\end{equation}
where
\begin{equation}
A^{\al}_{ij}(x)=\frac{1}{2}\pa_{[i}((\pa_m \theta^{\al})(\bx)
\si^m_{j]}(x)).
\end{equation}
Special co-ordinates $x^{i\prime}=(\theta^{\al}, y^\mu)$ can
always be found such that $\bx^{i\prime}(x^{\prime})=(0,y^\mu)$.
Note that the modified regularity condition \eqref{eq:Zsmooth}
implies $\theta^\al$ are adequate co-ordinates locally about any
point $x\in \M$. Then,
\begin{equation}
\om_{(p_\al,\theta^\ga ,y^\mu),(p_\be,\theta^\de,y^\nu)}=
\begin{pmatrix}
0 & \vdots & 1 & \vdots & 0\\
\hdotsfor{5}\\
-1 & \vdots & 0 & \vdots & 0\\
\hdotsfor{5}\\
0 & \vdots & 0 & \vdots & \om_{\mu\nu}(y)
\end{pmatrix}_{(p_\al,\theta^\ga ,y^\mu),(p_\be,\theta^\de,y^\nu)}.
\end{equation}
We can now see that if no non-degenerate co-ordinates $y^\mu$ on
the constraint surface $\Gamma$ can be found such that
$\om_{\mu\nu}(y)$ is constant, then $\om_{(p_\al,x^i),(p_\be,x^j)}$
isn't constant in any co-ordinate system. Thus, $S[\bx(x)]$ is only
`desirable' when canonical co-ordinates $y^\mu$ exist on the
constraint surface, this being one of the few examples where Dirac
bracket quantization is possible. In other words, we can quantize
no more systems using the BM form of the action $S[\bx(x)]$, than
by the standard Dirac bracket discussed in section
\ref{sec:Diracbracket}.

The repercussions of the above analysis appear on applying the
superfield version of the antifield (BV) formalism directly to the
BM action $S[\bx(x)]$, as performed in the subject paper
\cite{Batalin:2001hs}. One can directly write down the BRST
cohomology and after choosing a suitable gauge fixing fermion, jump
straight to the Fadeev-Popov style path integral. This Langranian
form of BRST quantization is exactly equivalent to following
through the above construction of $S_E[x,p_x]$ and the Hamiltonian
BRST (BFV) path integral and then integrating out the momenta,
leaving just $x^i$, Lagrange multipliers and ghosts.

First follows a short recap of this work:

The superfield version of $S[\bx(x)]$ is
\begin{equation}
S^{\prime}[\bx(x(t,\tau))]=\int{dt\,d\tau\,
\bx^i\bar{\om}_{ij}(\bx)D\bx^j} - \tau H(\bx),
\end{equation}
where $x^i(t,\tau)=x^i_0 (t) + \tau x^i_1 (t)$, $\tau$ is an odd
Grassmann parameter, $\eps(x^i_0)=\eps^i$, $\eps(x^i_1)=\eps^i+1$
and
\begin{equation}
D=\frac{d}{d\tau} + \tau\frac{d}{dt} \implies D^2=\frac{d}{dt}.
\end{equation}
The superfield master action is
\begin{equation}
\begin{split}
&S=S^{\prime}[\bx(x(t,\tau))] + \int{\tx_i \, d\tau\, dt\,
G^i_{\al}(x) C^\al}\\
&+ \frac{1}{2} \int{\tilde{C}_{\ga}\, d\tau \,
dt\, U^{\ga}_{\al\be}(x)C^{\be}C^{\al}(-1)^{\eps^\al}}
- \int{\tilde{\bar{C}}^{\al}\, d\tau\, dt\,
\la_{\al}(-1)^{\eps^{\al}}}
\end{split}
\end{equation}
where
\begin{eqnarray}\label{eq:master}
(S,S)=0 && \Delta S=0,
\end{eqnarray}
for superfields $\{x^i(t,\tau), C^\al(t,\tau), \bar{C}_\al(t,\tau),
\la_\al(t,\tau)\}$ and their super antifields. The superfield
measure has the remarkable property \cite{Batalin:1998pz}, of
behaving as a scalar under general, superfield co-ordinate
transformations, thus providing the natural measure. Equation
\eqref{eq:master} implies that no quantum corrections to the
measure are necessary.

The authors choose a satisfactory gauge fixing fermion which
determines the super antifields, leaving the gauge fixed action
\begin{equation}\label{eq:gfmaster}
\begin{split}
&S_{\psi}=S^{\prime}[\bx(x(t,\tau))] +
\int{d\tau\, dt\, \bar{C}_{\al} \theta^{\al}(x)\pl_i
G^i_{\be}(x)C^{\be}}\\
&- \int{d\tau\, dt\, \la_{\al} \theta^{\al}(x)}.
\end{split}
\end{equation}

There are various points to be made at this stage:

Elements of the zero ghost number cohomology $f(x)$ satisfy
\begin{equation}
(f(x),S)=0 \implies f(x) \pl_i G^i_{\al}(x)=0,
\end{equation}
where the ghosts play no role in closing the cohomological algebra.

Although the gauge fixed action \eqref{eq:gfmaster} is in standard
Fadeev-Popov form, with BRST symmetry
\begin{eqnarray}
\de x^i=\eta G^i_{\al}(x) C^{\al} && \de \la_{\al}=0\\
\de C^{\al}=-\frac{1}{2}\eta U^{\al}_{\be\ga} C^{\be}C^{\ga}  &&
\de\bar{C}_{\al}=\eta \la_{\al},
\end{eqnarray}
for Grassmann parameter $\eta$,  there's no associated conserved
Noether BRST charge. This follows from the fact that the gauge
symmetries of $S[\bx(x)]$ in the original phase space $\M$ have no
associated Noether charges. If we vary $x^i$ as $\de x^i =
G^i_{\al}(x) \eta^{\al}(t)$, where $\eta^{\al}$ is a small
parameter with Grassmann parity $\eps(\eta^{\al})=\eps^{\al}$, then
the Noether identity
\begin{equation}
\de S[\bx(x)]= \int{dt\, S[\bx]\frac{\dl}{\de \bx^k(t)}
(\bx^k\pl_i) G^i_{\al}(x)\eta^{\al}(t)}=0,
\end{equation}
disappears trivially. The equivalent gauge symmetries in $S_E[x,p]$
yield the first class constraints $\phi_{\al}$ as Noether charges.
These disappear when $p_{x^i}$ are integrated out of the path
integral.

The ghost part of the path integral disappears on integrating out
$C^{\al}$ and $\bar{C}_{\al}$, which seems to be associated
with the non-existence of a BRST operator. The identity
$\theta^{\al}(x)\pl_i G^i_{\be}(x)|_{\theta=0}=\de^{\al}_{\be}$
used in the subject paper is not needed to show this.

The end result is the superfield version \cite{Batalin:1998pz}, of
the standard, problematic 2nd class constraint path integral
\eqref{eq:xpathint}. So, in another way, we have come up against
the result that $S[\bx(x)]$ isn't a `desirable' action. This is
because, in parameterizing the 2nd class constraint surface
$\theta(x)=0$ degenerately, we haven't escaped from the fact that
we are still parameterizing the constraint surface with the same
problems as in non-degenerate parameterization, discussed in
section \ref{sec:Diracbracket}.

\section{Operator Quantization} \label{sec:operator}

As already alluded to in section \ref{sec:exploreBM}, the gauge
transformations \eqref{eq:gaugegen} have no corresponding Noether
charges within the phase space $\M$ and hence no generators.
Specifically, the Batalin Marnelius action $S[\bx(x)]$ isn't in
extended Hamiltonian form, as would be required for gauge
transformations to be generated by first class constraints.
However, to use a parameterized $S_E[x,p_x(x,p_\al)]$ with first
class constraint functions $\phi_{\al}=p_\al$ instead would be no
more advantageous, in that no more systems could be quantized than
using the approach below. Furthermore, $S_E[x,p_x(x,p_\al)]$ is
more cumbersome and, with 2M extra phase space co-ordinates
$p_\al$, not in the spirit of the subject paper
\cite{Batalin:2001hs}.

As we shall see, the following methods, are directly related to
the Dirac bracket approach in section \ref{sec:Diracbracket}.
Classically, all observables $F(x)$ satisfy
\begin{eqnarray}
F(x)\pl_k G^k_{\al}(x)=0 && \Imp F(x)=F(\bx(x))\\
\dot{F}(\bx)=\{F(\bx),H(\bx(x))\}_M,
\end{eqnarray}
where $\{,\}_M$ is the bracket defined in section
\ref{sec:Mybracket}.

There is a correspondence between the bracket $\{,\}_M$ on
degenerate co-ordinates $x^i$ and the bracket $\{,\}_*$ on
non-degenerate co-ordinates $y^\mu$. They have the similar
properties \eqref{eq:mybracket} and \eqref{eq:starbracket}, thus we
can write
\begin{equation}\label{eq:linkxy}
\{F(\bx(x)),G(\bx(x))\}_M= \{F(\bx(y)),G(\bx(y))\}_*
\end{equation}
for $\bx^i(x)=\bx^i(y)$. The only difference is that we only
allow gauge invariant functions, i.e. functions of the form
$F(\bx(x)$, inside the bracket $\{,\}_M$. All functions $f(y)$ can
be written in the form $F(\bx(y))$ however.

To emphasize this point, there is \textit{no} equivalent,
consistent, quantum gauge invariance condition on observables of
the form
\begin{equation}
[\hat{F}(x),\hat{G}_{\al}(x)]=V_{\al}^{\be} \hat{G}_\be,
\end{equation}
as there are no first class constraints within $\M$ associated
with $\pl_k G^k_{\al}(x)$. Thus, there is also \textit{no} gauge
invariance condition on the states $|s>$ in the Hilbert space of
the form
\begin{equation}
\hat{G}_{\al}|s>=0.
\end{equation}

One quantization procedure is as follows:

Firstly, we choose a basis for the classical observables
\begin{eqnarray}
Y^\mu(\bx(x)) && \mu=1,\ldots 2N-2M,
\end{eqnarray}
where $Y^\mu(\bx(x))=(y^\mu)^{-1}(\bx(x))$ and $y^\mu$ are
degenerate co-ordinates on the constraint surface $\Gamma$. So,
\begin{equation}
Y^\mu(\bx(y))=y^\mu.
\end{equation}
and
\begin{equation}
\begin{split}
\{Y^\mu(\bx(x)),Y^\nu(\bx(x))\}_M&=
\{Y^{\mu}(\bx(y),Y^{\nu}(\bx(y))\}_*\\
&=\om^{\mu\nu}(Y(\bx(x)))
\end{split}
\end{equation}
All observables $F(\bx(x))$ are written $f(Y(\bx(x)))$. We can
remove all mention of $\bx(x)$ and the problem reduces to finding a
representation of
\begin{equation}
[\Yh^\mu, \Yh^\nu]=i \hbar \om^{\mu\nu}(\Yh),
\end{equation}
just as in section \ref{sec:Diracbracket}.

A second procedure is as follows:

Sometimes, we can choose $\om_M^{ij}(x)$ such that a representation
of
\begin{equation}
[\xh^i,\xh^j]=i\hbar \om^{ij}_M(\xh)
\end{equation}
can be found. In particular, if $\om_M^{ij}=\om^{ij}$ and $x^i$ are
canonical co-ordinates (q,p), then we can construct the standard
co-ordinate representation. The states $|q>$, such that
$\hat{q^i}|q>=q^i|q>$ form a basis for the Hilbert space.

It's always possible to  make an appropriate choice of $\bx^i(x)$
and $\om_M^{ij}=\om^{ij}$ when there exist non-degenerate,
canonical co-ordinates $y^\mu$ on the constraint surface.

Clearly, unphysical operators and states are present in this
representation, and there's no way of projecting them out as in the
BRST formalism. By its very nature, all unphysical operators must
be removed at the classical stage.

So, we define some consistent normal ordering scheme, which
preserves the reality properties of observables
$:\hat{F}(\bx(\xh)):$. Now, for a co-ordinate representation, we
find $N-M$ commuting observables $:\Qh^{\ga}(\bx(\xh)):$
\begin{equation}
[:\Qh^{\ga}(\bx(\xh)):\, , \, :\Qh^{\de}(\bx(\xh)):]=0,
\end{equation}
then states $\{|Q>\}$, such that $\Qh^{\ga}|Q>=Q^{\ga}|Q>$, form a
basis for all physical states.

Essentially, this is the formalism advocated in
\cite{Lyakhovich:2001cm}, where the condition \eqref{eq:Poisson}
was actually imposed.

It's not always necessary to write down a basis for all physical
operators. We can consider a sub-algebra of operators, so long as we
can construct $:\Qh^{\ga}:$ from them in order to have a basis for
all states. Example \ref{exa:Ponsphere} illustrates this point, a
basis for the sub-algebra being $\{\Lh_i\}$ and the maximal set
of commuting operators $\{\Lhb^2, \Lh_3\}$.

\section{Mixed class constraints}

There are various ways that one can envisage introducing first
class constraints into this formalism. Supposing we start with the
usual irreducible second class constraints $\theta^{\al}(x)$ and
some irreducible first class constraints $g^a(x)$,

1. We could introduce good canonical gauge fixing constraints
\cite{Henneaux:1992ig} $c^a(x)$, such that $\text{sdet }
\{g^a(x),c^b(x)\} \neq 0$, then all the constraints are second
class and we can apply the BM formalism to this new system.

2. We could calculate $\bx^i(x)$ which parameterizes the second
class surface as before, however $g^a(\bx) \neq 0$ now. We then
quantize the action $S[\bx(x)]$. This was alluded to in
\cite{Lyakhovich:2001cm}.

Method 1 is unlikely to be desirable since first class constraints
are far easier to deal with than a BM processing of second class
ones, but it is method 2 which is most useful and which I shall
further explore below.

So after constructing $\bx^i(x)$ with the usual
properties \eqref{eq:xbar1} and \eqref{eq:xbar2}, one can write
down the action. The original extended action is
\begin{equation}
S_E[x]=\int{dt\, (x^i \bar{\om}_{ij}(x) \dot{x}^j
-H(x)-\lambda_{\al}\theta^{\al}(x) - \lambda_a g^a(x))}.
\end{equation}
The BM gauge invariant action is
\begin{equation}\label{eq:S_BM_firstclass}
S[\bx(x)]=\int{dt\, (\bx^i \bar{\om}_{ij}(\bx) \dot{\bx}^j -H(\bx)
-\lambda_a g^a(\bx))}.
\end{equation}
As well as the BM gauge symmetries from the degenerate
parameterization of $\bx(x)$ as denoted by
\begin{equation}
S[\bx]\frac{\dl}{\de x^i} G^i_{\al}(x)=0,
\end{equation}
there are gauge symmetries arising from $g^a(\bx)$. An
infinitesimal gauge transformation of a BM observable $F(\bx)$ is
\begin{equation}
\de_{\eta}F(\bx) =  \{F(\bx),\eta_a g^a(\bx)\}_M
\end{equation}
where $\eta_a$ are infinitesimal parameters with Grassmann parity
$\eps(\eta_a)=\eps(g^a(x))$. The relevant $\{,\}_M$ bracket
algebra is
\begin{eqnarray}
\{g_a(\bx),g_b(\bx)\}_M=U^c_{ab}(\bx)g_c(\bx) \\
\{H(\bx),g_a(\bx)\}_M=V_a^b(\bx)g_b(\bx)
\end{eqnarray}
and the gauge symmetry of the action $S[\bx]$ in
\eqref{eq:S_BM_firstclass} arising from $g_a$ is
\begin{eqnarray}
\de_{\eta} g_a(\bx)=\{g_a(\bx),\eta^b g_b(\bx)\}_M \\
\de_{\eta} H(\bx)=\{H(\bx), \eta^a g_a(\bx)\}_M \\
\de_{\eta} \la^a=\dot{\eta^a} + \la^c \eta^b U^a_{bc}(\bx) -\eta^b
V_b^a(\bx)
\end{eqnarray}
An observable $f(x)$ must be gauge invariant in both senses:
\begin{eqnarray}
f(x)\pl_i G^i_{\al}(x)=0 && \{f(x),g_a(\bx)\}_M=
W^b_a(\bx)g_a(\bx)
\end{eqnarray}
The above formalism is similar to that which we would obtain on
parameterizing the second class constraint surface with
non-degenerate co-ordinates $y^\mu$, as in section
\ref{sec:Diracbracket}. The approach is nearly identical to that
in \cite{Henneaux:1992ig}, with the bracket $\{,\}_*$ replaced by
$\{,\}_M$. The relation between the two brackets is evident in
equation \eqref{eq:linkxy}.

One can then either pursue the operator formalism as laid out in
section \ref{sec:operator}, or follow the superfield anti-field
approach, remembering the extra gauge symmetries from $g_a(\bx)$.

\section{Examples}
\exa[The simplest example]{
The simplest example that one can conceive is where there are
bosonic, canonical co-ordinates $\{x^i\}=\{q^1,q^2,p_1,p_2\}$ and
second class constraints
\begin{eqnarray}
\theta^1=q^1 && \theta^2=p_1.
\end{eqnarray}
For further simplicity, we don't consider the Hamiltonian, we just
generate the correct quantum representation of the system.

The simplest degenerate parameterization is
\begin{eqnarray}
\bq^1=0 && \bp_1=0\\
\bq^2=q^2 && \bp_2=p_2,
\end{eqnarray}
where
\begin{eqnarray}
Z^i_1 \pa_i=G^i_1 \pa_i=\frac{\pa}{\pa q^1} && Z^i_2 \pa_i=G^i_2
\pa_i=\frac{\pa}{\pa p_1}.
\end{eqnarray}
Now, $\bq^2$ and $\bp_2$ form a basis for the observables and in
this simple case,
\begin{equation}
\{\bx^i,\bx^j\}_M=\{\bx^i,\bx^j\}_D|_{x\ra \bx(x)}
\end{equation}
is solved by taking $\{,\}_M \equiv \{,\}$, where $\{,\}$ is the
standard Poisson bracket. So a basis of states is $\{|q^2>\}$ and
yields a representation of
\begin{eqnarray}
[\qh^2,\ph_2]=i\hbar && [\qh^2,\qh^2]=[\ph_2,\ph_2]=0.
\end{eqnarray}
}

\exa[A non-relativistic particle on a sphere]{

This example was both considered briefly in
\cite{Lyakhovich:2001cm}, and in detail in \cite{Harada:1990aj} in
the context of `gauge-unfixing'.

The action is
\begin{equation}
S=\int{dt\,(\frac{m}{2} \dot{\qb}^2 + \la (\qb^2-R^2))},
\end{equation}
where $\qb=(q^1,q^2,q^3)$, $R$ is the radius of the sphere and
$\la$ is a Lagrange multiplier. The Hamiltonian is
\begin{equation}
H=\frac{1}{2m}\pb^2 - \la(\qb^2-R^2),
\end{equation}
where $\pb=(p_1,p_2,p_3)$ and $p_\la$ are the canonical momenta
conjugate to $\qb$ and $\la$ respectively. The constraints are
\begin{eqnarray}
\theta^1=p_\la && \theta^2= \qb^2-R^2\\
\theta^3=\qb .\pb && \theta^4=\la -\frac{\pb^2}{2mR^2}.
\end{eqnarray}
where $\theta^2,\theta^3,\theta^4$ come from requiring the
consistency conditions $\dot{\theta^\al}\approx 0$. We can
calculate $\det{\{\theta^\al,\theta^\be\}}\neq 0$ so long as $\qb^2
\neq 0$, thus the constraints are second class.

I now compare various different methods.

The `gauge unfixing' procedure, discussed in \cite{Harada:1990aj},
was to discard the constraints $\theta^3,\theta^4$, leaving first
class constraints $\theta^1$ and $\theta^2$.
\begin{equation}
\tilde{H}=\frac{\Lb^2}{2mR^2}
\end{equation}
is then a suitable first class Hamiltonian i.e. $\{\tilde{H},
\theta^\al\}|_{\theta^1=\theta^2=0}=0$ for $\al=1,2$, with
$L_i=\eps_{ijk}q^jp_k$ and
\begin{equation}\label{eq:Lalg}
\{L_i,L_j\}=\eps_{ijk}L_k
\end{equation}
as the standard angular momenta algebra.

The total Hilbert space is spanned by $|\la; r; \theta,\phi>$,
where $r^2=\qb^2$ and $\theta,\phi$ are spherical polar
co-ordinates. A basis for physical states is $|p_\la=0;r=R;l,m>$,
from which a consistent representation of sub-algebra
\begin{equation}\label{eq:QLalg}
[\Lh_i,\Lh_j]=i\hbar\Lh_k
\end{equation}
is formed. The states $|l,m>$ are labelled, by employing commuting
operators $\Lhb^2$ and $\Lh_3$, as follows:
\begin{eqnarray}
\Lhb^2|l,m>=\hbar^2 l(l+1) && \Lh_3|l,m>=\hbar m|l,m>.
\end{eqnarray}
The physical states are energy eigenstates, with eigenvalues
\begin{equation}
E_l=\frac{\hbar^2 l(l+1)}{2mR^2}.
\end{equation}
The point here, is that we use the standard Poisson bracket from
which a quantum representation is always possible.

A second method is the non-degenerate parameterization technique
, discussed in section \ref{sec:Diracbracket}.
We can take spherical polar co-ordinates such that
\begin{eqnarray}
\bar{\qb}=R\erb && \bar{\pb}=\frac{p_{\theta}}{R}\etb +
\frac{p_{\phi}}{R\sin{\theta}}\epb
\end{eqnarray}
where
\begin{eqnarray}
\erb=
\begin{pmatrix}
\sin{\theta}\cos{\phi}\\
\sin{\theta}\sin{\phi}\\
\cos{\theta}
\end{pmatrix}&&
\etb=
\begin{pmatrix}
\cos{\theta}\cos{\phi}\\
\cos{\theta}\sin{\phi}\\
-\sin{\theta}
\end{pmatrix}\\
\epb=
\begin{pmatrix}
-\sin{\phi}\\
\cos{\phi}\\
0
\end{pmatrix}.
\end{eqnarray}
Constraints $\theta^1$ and $\theta^4$ are simply parameterized away
using their defining equations. Applying equation \eqref{eq:omy},
one can show that $\{\theta,\phi,p_{\theta},p_{\phi}\}$ are
canonical co-ordinates on the constraint surface, thus a quantum
representation can be found. Writing
$\Lb=\bar{\qb}\times\bar{\pb}$, we find that the Hamiltonian is
\begin{equation}
H=\frac{\Lb^2(\bar{\qb},\bar{\pb})}{2mR^2}
\end{equation}
and using $\{\theta, p_{\theta}\}_*=1$ etc. ,
\begin{equation}
\{L_i,L_j\}_*=\eps_{ijk}L_k.
\end{equation}
A quantum representation of \eqref{eq:QLalg} is constructed.
The full Hilbert space of states is spanned by $\{|l,m>\}$, where
$\Lhb^2$ and $\Lh_3$ form the maximal set of commuting operators.

A third method is the second operator formalism of section
\ref{sec:operator}, which in this case, is the same as that in
\cite{Lyakhovich:2001cm}.

A degenerate parameterization is
\begin{eqnarray}
\bar{\qb}=\frac{R}{\sqrt{\qb^2}}\qb &&
\bar{\pb}=\frac{\sqrt{\qb^2}}{R}(\pb-\frac{\pb.\qb}{\qb^2}\qb),
\end{eqnarray}
where we can see that conditions \eqref{eq:xbar1} and
\eqref{eq:map2} are satisfied.
The Hamiltonian is
\begin{equation}
H=\frac{\bar{\pb}^2}{2m}= \frac{\Lb^2(\qb,\pb)}{2mR^2}.
\end{equation}
We can check that $\{,\}_M \equiv \{,\}$ in this case, and also
that $\Lb(\bar{\qb},\bar{\pb})=\Lb(\qb,\pb)$. Thus the functions
$L_i$ are observables and form the closed algebra as in
\eqref{eq:Lalg}. We can form a quantum representation of the
sub-algebra \eqref{eq:QLalg}, and again label states $|l,m>$ with
the maximal commuting sub-algebra of observables, $\Lhb^2$ and
$\Lh_3$.

The point was made in \cite{Lyakhovich:2001cm} of the similarity
between the above degenerate parameterization quantization
approach, and the `gauge-unfixing' procedure in the first part of
this example. As I have argued in sections \ref{sec:exploreBM} and
\ref{sec:operator} of this paper, the similarities between the
degenerate and non-degenerate parameterization are more
fundamental, being linked by equation \eqref{eq:linkxy}.
}

\label{exa:Ponsphere}

\section{Concluding remarks}

In the preceding work, it was shown that the regularity conditions
on the constraints needed to be modified, and also, the bracket
$\{,\}_M$ was introduced from which was developed the operator
formalism. We have seen that quantization of the BM theory amounts
to Dirac bracket quantization, and lacks the power of other
established approaches
\cite{Batalin:1987fm,Batalin:1990mb,Batalin:1991jm} and
\cite{Harada:1990aj,Mitra:1990mp} mentioned in the introduction.

However, where Dirac bracket quantization is possible, the second
of the operator formalisms in section \ref{sec:operator} is an
attractive alternative to the methods used in section
\ref{sec:Diracbracket}. It combines the advantage, as with
non-degenerate parameterization, of imposing the constraints at
the classical stage whilst avoiding the, often inconvenient,
co-ordinates $y^\mu$. Example \ref{exa:Ponsphere}, `a
non-relativistic particle on a sphere', illustrates this point.

\medskip
\section*{Acknowledgments}
I am grateful to Oleg Soloviev for his support during my PhD. I
would also like to thank Igor Batalin, Marc Henneaux, Robert
Marnelius and Daniel Waldram for answering my questions.

This work was funded by PPARC.

-----------------------------------------------------------------
\bibliographystyle{utphys}
\bibliography{xbib}


\end{document}